\documentclass[twocolumn,prb,superscriptaddress,longbibliography]{revtex4-2}
\usepackage{bm}
\usepackage{amssymb,amsmath}
\usepackage{graphicx}
\usepackage{color}
\usepackage[colorlinks=true,urlcolor=blue,citecolor=blue]{hyperref}

\begin{document}
\title{Magnetic properties of diluted hexaferrites}
\author{Logan Sowadski}
\affiliation{Department of Physics, Missouri University of Science and Technology, Rolla, MO 65409, USA}
\author{Sean Anderson}
\affiliation{Department of Physics, Missouri University of Science and Technology, Rolla, MO 65409, USA}
\author{Cameron Lerch}
\affiliation{Department of Mechanical Engineering and Materials Science, Yale University, New Haven, Connecticut 06520, USA}
\author{Julia Medvedeva}
\affiliation{Department of Physics, Missouri University of Science and Technology, Rolla, MO 65409, USA}
\author{Thomas Vojta}
\affiliation{Department of Physics, Missouri University of Science and Technology, Rolla, MO 65409, USA}

\begin{abstract}
We revisit the magnetic properties of the hexagonal ferrite PbFe$_{12-x}$Ga$_x$O$_{19}$. Recent experiments have
reported puzzling dependencies of the ordering temperature and the saturation magnetization on the Ga concentration $x$.
To explain these observations, we perform large-scale Monte Carlo simulations, focusing on the effects of an unequal
distribution of the Ga impurities over the five distinct Fe sublattices. Ab-initio density-functional calculations
predict that the Ga ions preferably occupy the $12k$ sublattice and (to a lesser extent) the $2a$ sublattice. We incorporate
this insight into a nonuniform model of the Ga distribution. Monte Carlo simulations using this model lead to an
excellent agreement between the theoretical and experimental values of the ordering temperature and saturation magnetization,
indicating that the unequal distribution of the Ga impurities is the main reason for the unusual magnetic properties of PbFe$_{12-x}$Ga$_x$O$_{19}$.
We also compute the temperature and concentration dependencies of the sublattice magnetizations, and we study the character
of the zero-temperature transition that takes place when the ordering temperature is tuned to zero.
\end{abstract}

\date{\today}

\maketitle

\section{Introduction}
\label{sec:intro}

Recent years have seen renewed interest in the properties of hexagonal ferrites (hexaferrites). These materials have
numerous technological applications including permanent magnets, magnetic recording and data storage devices,
as well as high-frequency electronics \cite{Pullar12,Pullar14}. In addition, they feature interesting magnetic and
ferroelectric quantum behavior at low temperatures \cite{Rowleyetal16,Shenetal16,Rowleyetal17}.

\begin{figure}[b]
\includegraphics[width=\columnwidth]{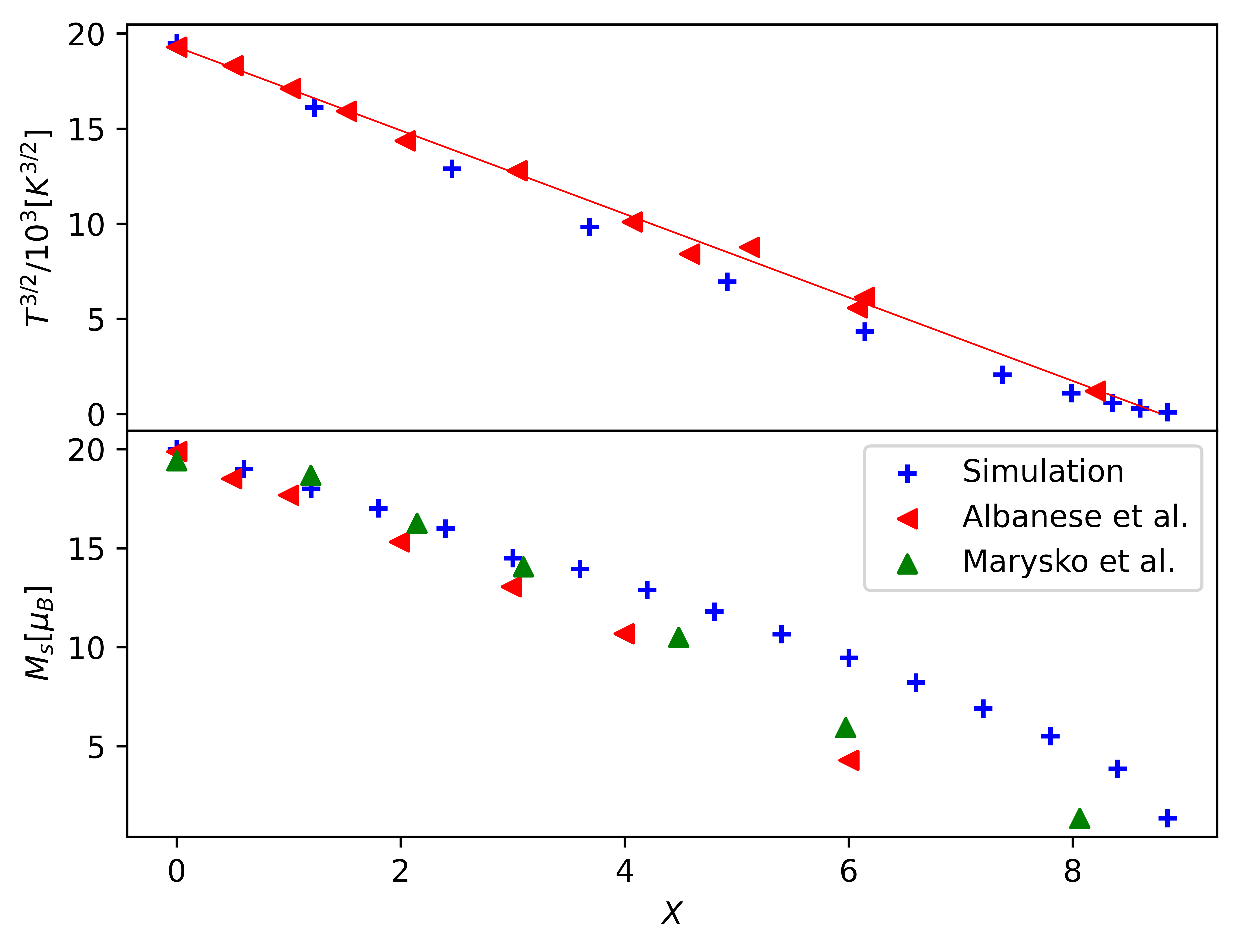}
\caption{Magnetic ordering temperature $T_c$ (top) and low-temperature saturation magnetization $M_s$
per formula unit (bottom) of PbFe$_{12-x}$Ga$_x$O$_{19}$
as functions of the Ga concentration $x$. The experimental data for $T_c$ are taken from Refs.\ \cite{Rowleyetal17,ALWD02}, and the
$M_s$ data stem from Refs.\ \cite{MaryskoFraitKrupicka97,ALWD02}. The solid line represents the 2/3-power law for $T_c$ put
forward in Ref.\ \cite{Rowleyetal17}. The Monte Carlo simulations assumed a uniform distribution of the Ga atoms
over all available iron sites. The results for $T_c$ are from Ref.\ \cite{KhairnarLerchVojta21} whereas
those for $M_s$ were computed here, using the algorithm of Ref.\ \cite{KhairnarLerchVojta21}.}
\label{fig:experiment}
\end{figure}

The magnetic properties of hexaferrites can be tuned by diluting the magnetic degrees of freedom. Several
experimental studies \cite{MaryskoFraitKrupicka97,ALWD02,Rowleyetal17} reported the results of randomly
substituting nonmagnetic Ga ions for the magnetic Fe ions in magnetoplumbite, PbFe$_{12}$O$_{19}$.
Magnetoplumbite is a Lieb-Mattis type ferrimagnet \cite{LiebMattis62} with a magnetic ordering
temperature $T_c$ of about 720 K and a
low-temperature saturation magnetization $M_s$ of 20$\mu_B$ per formula unit. The magnetic ordering
temperature of PbFe$_{12-x}$Ga$_x$O$_{19}$ decreases with increasing Ga concentration $x$
and vanishes at a critical concentration $x_c \approx 8.6$. The value of $x_c$ is very close to the site percolation
threshold of the lattice spanned by the exchange interactions between the Fe ions, suggesting that the zero-temperature
magnetic phase transition at $x_c$ is of percolation type \cite{Rowleyetal17}. The magnetic phase boundary can be
approximated well by the relation $T_c(x) = T_c(0) (1-x/x_c$)$^\phi$ with $\phi = 2/3$ over the entire concentration
range. Interestingly, the low-temperature saturation magnetization $M_s$ decreases much faster with $x$ than $T_c$,
as is shown in Fig.\ \ref{fig:experiment}.

To explain these findings, Khairnar et al.\ \cite{KhairnarLerchVojta21} performed Monte Carlo simulations of a randomly diluted
Heisenberg model, employing the magnetoplumbite crystal structure and realistic exchange interactions. As shown in Fig.\
\ref{fig:experiment}, the results of these simulations did not agree with the experimental data. Specifically, the ordering temperature predicted by the simulations is \emph{lower} than the experimental values and does not follow the striking 2/3
power law, whereas the saturation magnetization predicted by the simulations is significantly \emph{higher} than the experimental findings.
This is a puzzling situation for at least two reasons. First, while it is relatively easy to identify possible mechanisms that
could lead to a faster reduction of $T_c$ with $x$ compared to a simple Heisenberg Hamiltonian (e.g., frustrating interactions, non-collinear order, or quantum fluctuations) it is harder to find reasons for the experimental $T_c$ to decay more slowly than
the model calculation. Second, one would usually assume
that a mechanism that increases $T_c$ to also increase $M_s$, but the experimental $M_s$ values are significantly below the simulation results. In summary, these finding imply that our understanding of the magnetic properties of the diluted hexaferrites remains incomplete, especially at higher dilutions.

\begin{figure}
\centerline{\includegraphics[width=7cm]{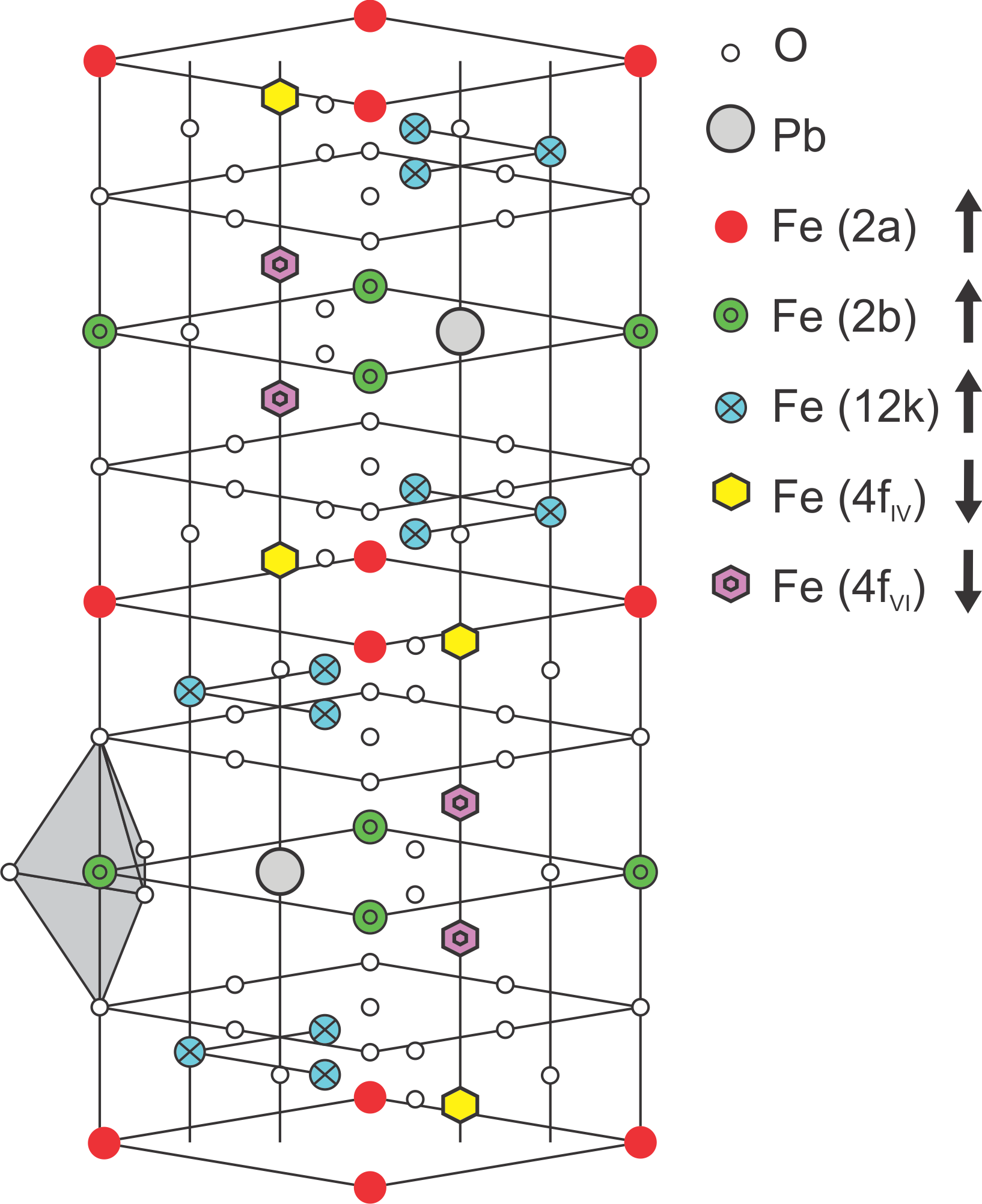}}
\caption{Double unit cell of PbFe$_{12}$O$_{19}$. Twelve Fe$^{3+}$ ions per unit cell are located on five distinct sublattices.}
\label{fig:crystal}
\end{figure}

The $M$-type hexaferrites PbFe$_{12}$O$_{19}$, BaFe$_{12}$O$_{19}$ and SrFe$_{12}$O$_{19}$ crystalize in the
magnetoplumbite structure presented in Fig.\ \ref{fig:crystal}.
The twelve Fe$^{3+}$ ions per unit cell,
each in the S=5/2 spin state,  are located on five distinct sublattices: 6 ions on the octahedral $12k$ sublattice,
one ion on the octahedral $2a$ sublattice, one ion on the pseudohexahedral $2b$ sublattice, two ions on the tetrahedral $4f_\mathrm{IV}$
sublattice, and two ions on the octahedral $4f_\mathrm{VI}$ sublattice.
Below $T_c$ of about 720 K, the spins feature collinear ferrimagnetic order with eight spins ($12k$, $2a$, and $2b$) pointing up and four spins
($4f_\mathrm{IV}$ and $4f_\mathrm{VI}$) pointing down.

As the crystal structure contains five inequivalent iron sublattices,
the distribution of the gallium ions over these sublattices
can be expected to play an important role for the magnetic properties. In fact, this question has been considered in several
publications in the literature, with inconclusive results. Marysko et al.\ \cite{MaryskoFraitKrupicka97} concluded from their
magnetic measurements and ferromagnetic resonance experiments that
the Ga$^{3+}$ ions are distributed over all sublattices except the $2b$ sublattice, at least for $x$ up to about 4
(in analogy with earlier results for other hexaferrites \cite{AlbaneseCarbucicchioDeriu73,AlbaneseCarbucicchioDeriu74}).
In contrast, Albanese et al.\ \cite{ALWD02} more recently reported accurate M\"ossbauer measurements indicating that the Ga$^{3+}$
ions are distributed over all five sublattices with nearly equal probability (even though a slightly higher gallium
concentration in the spin-up sublattices could not be excluded). For the related compound SrFe$_{12-x}$Ga$_x$O$_{19}$
(whose $T_c(x)$ curve is virtually indistinguishable from that of PbFe$_{12-x}$Ga$_x$O$_{19}$ \cite{ALWD02}),
M\"ossbauer studies \cite{ThompsonEvans94} suggested that the Ga ions preferably occupy the octahedral $4f_\mathrm{VI}$ site.
First-principle calculations \cite{Dixitetal15}, in contrast, show a strong preference of the Ga ions for the $12k$ sublattice.

The percolation calculations in Ref.\ \cite{Rowleyetal17} as well as the Monte Carlo simulations of
Ref.\ \cite{KhairnarLerchVojta21} were performed under the assumption that the Ga impurities are distributed
with equal probability over all sublattices.
In view of the disagreement between the magnetic measurements on PbFe$_{12-x}$Ga$_x$O$_{19}$ and the results of the
Monte Carlo simulations in the literature, it is prudent to revisit this assumption.

In the present paper, we therefore combine ab-initio density functional calculations and large-scale Monte Carlo simulations
to systematically study how an unequal gallium distribution over the available sublattices affects the magnetic properties
of PbFe$_{12-x}$Ga$_x$O$_{19}$. Our results can be summarized as follows. According to the density functional calculations,
the $12k$ sublattice is the most favorable location for the Ga$^{3+}$ ions, followed by the $2a$ sublattice. Ga$^{3+}$ ions
in any of the other sublattices lead to significantly higher total energies. We use this insight to construct a diluted
Heisenberg Hamiltonian with a biased
distribution of spinless impurities. Monte Carlo simulations of this Hamiltonian lead to an excellent agreement between the
experimental data for the ordering temperature $T_c$ and the low-temperature saturation magnetization $M_s$ and the
corresponding simulation results. This indicates that the unequal distribution of gallium impurities is likely the main reason
for the unusual magnetic behavior of PbFe$_{12-x}$Ga$_x$O$_{19}$.

The rest of the paper is organized as follows. In Sec.\ \ref{sec:model}, we introduce the site-diluted Heisenberg Hamiltonian,
and we discuss the density functional calculations that inform our model of the impurity distribution. The Monte Carlo
simulation methods and data analysis techniques are described in Sec.\ \ref{sec:MC}. Section \ref{sec:results} is devoted
to the simulation results and the comparison with the experimental data. We conclude in Sec.\ \ref{sec:conclusions}.

\section{Model}
\label{sec:model}
\subsection{Site-diluted Heisenberg Hamiltonian}
\label{subsec:heisenberg}

The high spin value $S=5/2$ of the Fe$^{3+}$ ions and the high ordering temperature of about 720 K for the undiluted
compound suggest that a classical approach to the magnetic degrees of freedom should provide a good
approximation. To describe the magnetism of PbFe$_{12-x}$Ga$_x$O$_{19}$, we therefore define a classical Heisenberg
model by placing either a classical Heisenberg spin or a vacancy on each Fe site in the hexaferrite crystal structure.
The Hamiltonian is given by
\begin{equation}
H = \sum_{i,j} J_{ij} \epsilon_i \epsilon_j \mathbf{S}_i \mathbf{S}_j~.
\label{eq:H_hexa}
\end{equation}
Here, $\mathbf{S}_i$ is an $O(3)$ unit vector at site $i$. The exchange interactions $J_{ij}$ are all positive, i.e.,
antiferromagnetic. We base their values on density functional calculations in Refs.\ \cite{NovakRusz05,Wuetal16}
(for BaFe$_{12}$O$_{19}$) and Ref.\ \cite{TejeraCentenoGallegoCerda21} (for SrFe$_{12}$O$_{19}$).
The resulting interactions depend weakly on the value of the effective Hubbard interaction energy $U_\textrm{eff}$ assumed in the density functional algorithm,
and they vary somewhat between the different calculations. However, scaling the interactions
by a common factor to reproduce the clean ordering temperature $T_c =720K$ of the undiluted material
suppresses most of these variations.
In our simulations, we only include the strongest interactions which are between the following sublattice pairs: $2a-4f_\mathrm{IV}$,  $2b-4f_\mathrm{VI}$, $12k-4f_\mathrm{IV}$, $12k-4f_\mathrm{VI}$. Their values are listed in Table \ref{table:J}.
\begin{table}
\begin{tabular*}{\columnwidth}{@{\extracolsep{\fill}}lcccc@{}}
\hline\hline
sublattice pair & $2a-4f_\mathrm{IV}$ &  $2b-4f_\mathrm{VI}$ & $12k-4f_\mathrm{IV}$ & $12k-4f_\mathrm{VI}$ \\
\hline
$K_{ij}$         &  5 meV       &  5 meV       &  3.5 meV      &    5 meV \\
$J_{ij}$         & 439  K       & 439  K       & 311 K         &   439 K      \\
\hline\hline
\end{tabular*}
\caption{Values of the exchange interactions. $K_{ij}$ denotes the interactions computed in
Ref.\ \cite{Wuetal16} for an effective Hubbard interaction $U_\mathrm{eff}=6.7$. We have absorbed the values of the magnetic moments into the interactions
$J_{ij}$ used in the Heisenberg Hamiltonian (\ref{eq:H_hexa}) and scaled them with a constant factor
$c=1.212$ to reproduce the clean ordering temperature of 720 K, i.e., $J_{ij} = (5/2)^2 c K_{ij}$}
\label{table:J}
\end{table}
These interactions are non-frustrated and establish the ferrimagnetic order.
We will discuss in the concluding section the effects of additional couplings which are significantly
weaker but frustrate the ferrimagnetic order.
We employ the exchange interactions computed for the undiluted system for our simulations in entire $x$-range.
This neglects variations of the interactions caused by changes in the lattice geometry due the substitution of
Fe ions by Ga ions. These changes are expected to be small because of the small difference
between the ionic radii of Ga$^{3+}$ (0.62 \AA) and Fe$^{3+}$ (0.64 \AA) cations \cite{Shannon76}.

The $\epsilon_i$ are independent quenched random variables that implement the site dilution. The can take the values
0 (vacancy) with probability $p_i$ and 1 (occupied site) with probability $1-p_i$. In the simulations performed in Ref.\ \cite{KhairnarLerchVojta21}, all lattice sites were assumed to feature the same vacancy probabilities, $p_i = p$ which
is related to the average number of Ga ions in the unit cell via $p = x/12$. The goal of the present paper is
to explore the effects of deviations from such a uniform Ga distribution. Nonuniform Ga distributions
are discussed in detail in the next subsection.

\subsection{Distribution of the Ga impurities}
\label{subsec:Gadistribution}

As pointed out in Sec.\ \ref{sec:intro}, the available results on the distributions of the Ga ions over
the five Fe sublattices \cite{MaryskoFraitKrupicka97,ALWD02} are inconclusive and partially contradict
each other. We therefore perform state-of-the-art ab-initio density functional calculations to determine the
preferred locations of the Ga impurities (for details see Appendix A).

\begin{figure}
\includegraphics[width=\columnwidth]{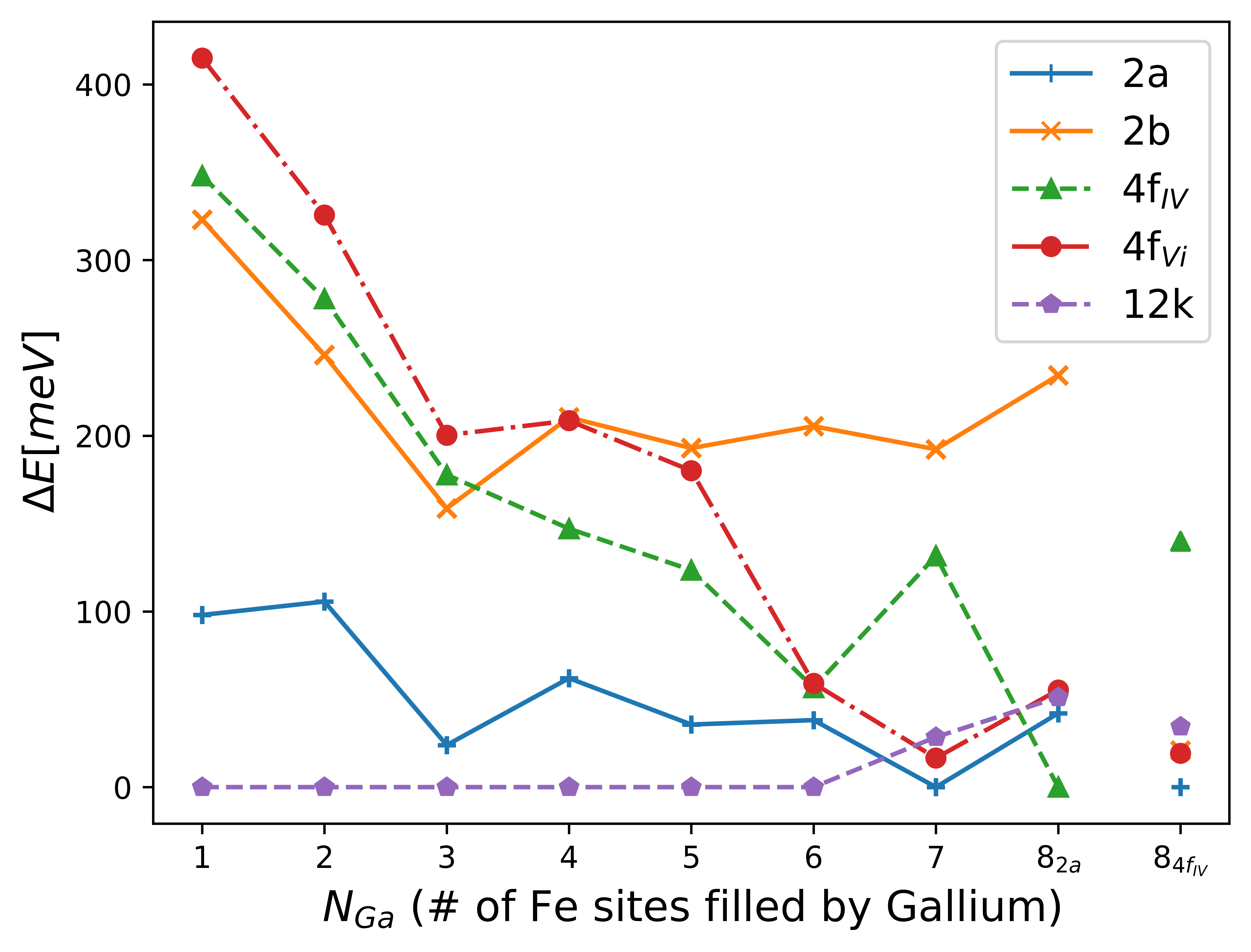}
\caption{Energy cost $\Delta E$ of the $N$-th Ga impurity in a PbFe$_{12}$O$_{19}$ double unit cell relative to the most favorable position
in the double unit cell. The energy of the $N$-th Ga impurity is computed while fixing Ga impurities 1 to $N-1$ at their most favorable
position (see text for details). We test two possible positions of the 7th impurity ($2a$ and $4f_\mathrm{IV}$) when calculating the energy cost of the 8th one.}
\label{fig:dft2}
\end{figure}

Specifically, we calculate the energies $E_{m}$ required for the Ga atoms to occupy different Fe sites in the
lattice. This is done in an iterative way. We start from an undiluted double unit cell and compute the energy of a
single Ga substitution in one of the five sublattices. As shown in Fig.\ \ref{fig:dft2}, we find that the first Ga impurity prefers the
12k sublattice by about 0.1\,eV over the 2a sublattice. All other sublattices have much higher energies.
For more than one impurity, we employ an iterative procedure. We fix the positions of Ga atoms 1 to $N-1$ at their most favorable positions and then
compute the energies for all possible positions of the $N$-th Ga atom. Only the lowest energy configurations are shown in Fig.\ \ref{fig:dft2}
for the $N$-th Ga atom among all considered non-equivalent positions with respect to the $N$-1 Ga substitutions.
We find that at low dilutions, the Ga impurities overwhelmingly favor filling sites in the 12k sublattice, followed by the 2a sublattice.
It should be noted that the energy difference between the two sublattices decreases to below 45 meV at $N$ = 3, 5 and 6.
In addition, the energy difference with both $4f$ sublattices decreases steadily as the Ga concentration at the 12k sites increases.
Beyond a dilution of about $p = 0.25$ (6 impurities in the double unit cell), the 12k preference changes,
and several sublattices have comparable $E_m$, suggesting a more even distribution of the impurities.
We note that in the lowest energy configurations, the Fe-O distances and the Fe magnetic moments change remarkably little during
the dilution process among all Fe sublattices except for the 2b, the pseudohexahedral site. The results are consistent with the 4- and 6-fold
 Ga coordination with oxygen atoms in monoclinic $\beta$-Ga$_2$O$_3$ structure. This is despite the sizable changes in mass and magnetization of our system
(with the largest increase of 5\% in the negative magnetic moments of Fe $4f$ sites).

To model a nonuniform vacancy distribution in the Heisenberg Hamiltonian (\ref{eq:H_hexa}), we introduce
weights $w(2a)$,  $w(2b)$, $w(4f_\mathrm{IV})$, $w(4f_\mathrm{VI})$, and $w(12k)$ that modify the vacancy probability
compared to the uniform case. Specifically, the vacancy probability in sublattice $Y$ is given by $w(Y) \, p$.
(This means that the uniform case is recovered if all weights are equal to unity.)
In the presence of these weights, the average number of vacancies in a unit cell is given by $w(2a) \, p + w(2b) \, p + 2 w(4f_\mathrm{IV}) \, p +
2 w(4f_\mathrm{VI}) \, p + 6 w(12k) \, p$. This implies that the weights have to fulfill the constraint
\begin{equation}
w(2a) + w(2b) + 2 w(4f_\mathrm{IV}) + 2 w(4f_\mathrm{VI}) + 6 w(12k) =12
\label{eq:constraint}
\end{equation}
to ensure that the overall vacancy probability in the system still equals $p$ and, equivalently, the average number of vacancies per unit cell is 12$p$

Motivated by the density functional results, we focus on model distributions where $w(12k)$ is larger than unity
while all other weights are identical to each other and smaller than unity. Inserting $w(2a)=w(2b)=w(4f_\mathrm{IV})=w(4f_\mathrm{VI})$ into the constraint (\ref{eq:constraint})
yields
\begin{equation}
w(2a)=w(2b)=w(4f_\mathrm{IV})=w(4f_\mathrm{VI})=2-w(12k) ~.
\label{eq:w12k_only}
\end{equation}
We also perform a few exploratory
calculations for models in which both $w(12k)$ and $w(2a)$ are increased while all others weights are identical
and decreased compared to unity.

\section{Monte Carlo simulations}
\label{sec:MC}
\subsection{Algorithm}
\label{subsec:algorithm}

We carry out large-scale Monte Carlo simulations of the classical Heisenberg model (\ref{eq:H_hexa}) to determine the
magnetic ordering temperature $T_c$, the saturation magnetization $M_s$, and other magnetic quantities.
These simulations utilize both Wolff cluster updates \cite{Wolff89} and Metropolis single-spin updates \cite{MRRT53}.
The Wolff algorithm greatly reduces the critical slowing down of the system near criticality and thus allows us to study
large systems with reasonable numerical effort. However, the Wolff algorithm alone is not sufficient in the presence of
site dilution because small isolated clusters of spins may form that are not connected to the bulk of the system. The Wolff cluster
construction cannot reach such isolated spins and therefore fails to equilibrate them. To overcome this problem, we complement the Wolff algorithm
with single-spin Metropolis updates which consider all spins including those that are isolated from the bulk.
Specifically, a full Monte Carlo sweep in our simulations consists of a Wolff sweep (a number of cluster flips such that the
total number of flipped spins equals the number of Fe lattice sites in the system)  followed by a Metropolis sweep (one attempted single-spin
flip per lattice site).

We simulate systems consisting of  $L^3$ double unit cells with $L$ ranging from 6 to 48. As each double unit cell contains
$24$ Fe sites, our largest systems contain about 2.6 million spins. All physical quantities of interest are averaged over
6400 to 25,600 independent disorder configurations for each size. Statistical errors are obtained from the variations
of the results between the configurations.

\begin{figure}
\includegraphics[width=\columnwidth]{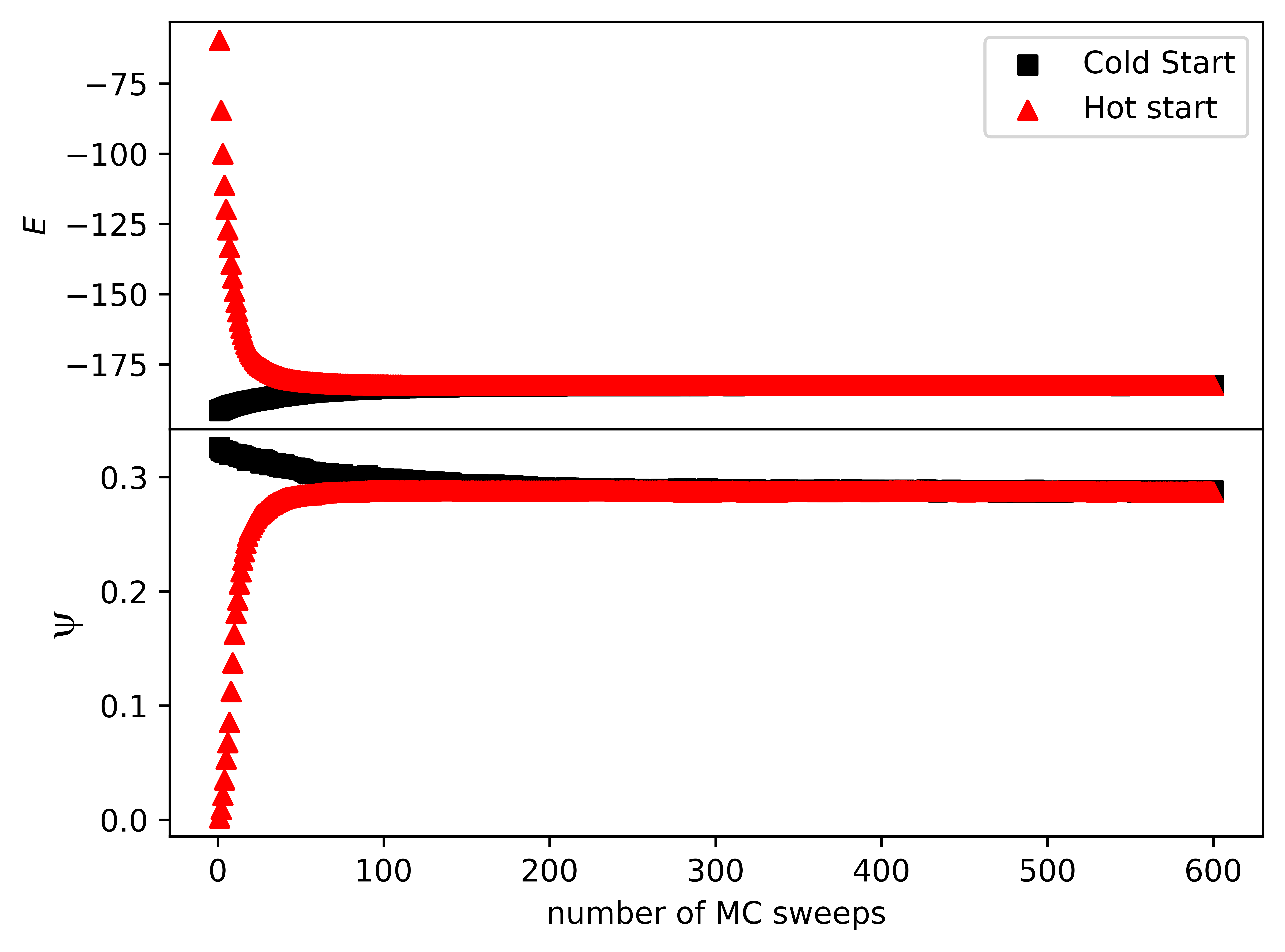}
\caption{Equilibration of the energy $E$ per site and the order parameter $\psi$ for a system of $48^3$ double unit cells, dilution
$p=0.66$, weight $w(12k)=1.25$ and temperature $T=27$\,K. The data are averages over 10 runs. The comparison of hot and cold starts
shows that the
system equilibrates after roughly 300 Monte Carlo sweeps despite being close to the critical point.}
\label{fig:eqsweeps}
\end{figure}

To find the number of Monte Carlo sweeps required for the system to equilibrate, we compare the results of
runs with hot starts (for which the spins initially point in random directions) and with cold starts (for which all spins are
initially aligned with the ferrimagnetic order). An example of such a test for a system close to its critical point is shown in
Fig. \ref{fig:eqsweeps}.
The energy and order parameter reach their equilibrium values after roughly 300 Monte Carlo sweeps.
Similar numerical checks were performed for other parameter values. Based on these
tests, we have chosen to perform 1000 equilibration sweeps and 2000 measurement sweeps.
Note that performing comparatively short Monte Carlo runs for a large number of disorder configurations
reduces the total statistical error \cite{BFMM98,VojtaSknepnek06,ZWNHV15}.

\subsection{Data analysis}
\label{subsec:dataanlysis}

The order parameter $\bm{\psi}$ of the ferrimagnetic transition in the hexaferrites is the ``staggered'' magnetization that counts
the spin-up sublattices positive and the spin-down sublattices negative,
\begin{equation}
\bm{\psi} = \frac 1 N \sum_i f_i \epsilon_i \mathbf{S}_i
\label{eq:OP}
\end{equation}
where $N$ is the total number of lattices sites. $f_i=1$ for sites in the $2a$, $2b$, and $12k$ sublattices whereas $f_i=-1$ for
the $4f_\mathrm{IV}$ and $4f_\mathrm{VI}$ sublattices. In contrast, the physical magnetization is given by
\begin{equation}
\mathbf{m} = \frac 1 N \sum_i \epsilon_i \mathbf{S}_i.
\label{eq:mag}
\end{equation}
For easier comparison with experiment, we will use $\mathbf{M} = 20 \mu_B \, \mathbf{m}$, which specifies the magnetization in
Bohr magnetons per formula unit. We also compute the sublattice magnetizations $\mathbf{m}_l$ for each of the five Fe sublattices.
They are defined as
\begin{equation}
\mathbf{m}_l = \frac 1 {N_l} \sum_{i\in l} \epsilon_i \mathbf{S}_i
\label{eq:ml}
\end{equation}
where the sum runs over all sites in the sublattice, and $N_l$ is their number.

To determine the ordering temperature $T_c$, we analyze the crossing of the Binder cumulant vs.\ temperature curves \cite{Binder81}.
The Binder cumulant of the order parameter is defined as
\begin{equation}
g = \left[1-\frac{\langle|\bm{\psi}|^4\rangle}{3\langle|\bm{\psi}|^2\rangle^2}\right]_{dis}~.
\end{equation}
Here $\langle...\rangle$ denotes the thermodynamic (Monte Carlo) average and $[...]_{dis}$ denotes the average
over disorder configurations. The behavior of the Binder cumulant in the thermodynamic limit of infinite
system size is well understood. In the ordered phase, $T < T_c$, all spins are correlated, and the magnetization has negligible
fluctuations around a nonzero value. Therefore, $\langle|\bm{\psi}|^4 \rangle \approx \langle|\bm{\psi}|^2\rangle^2$, and the
Binder cumulant approaches 2/3. In the disordered phase, $T > T_c$, the system consists of many
independent fluctuators. Consequently, $\langle|\bm{\psi}|^4 \rangle$ can be decomposed using Wick's theorem.
For O(3) symmetry this gives $\langle|\bm{\psi}|^4 \rangle \approx (15/9) \langle|\bm{\psi}|^2\rangle^2$, and the Binder
cumulant approaches 4/9. For finite system size, this step-function dependence of $g$ on temperature is rounded.

Because the Binder cumulant $g$ is a dimensionless quantity, it fulfills the finite-size scaling \cite{Barber_review83} form
\begin{equation}
g(t,L,u) = g(t\lambda^{-1/\nu},L\lambda,u\lambda^\delta)~.
\label{eq:g_FSS}
\end{equation}
Here, $\lambda$ is an arbitrary length scale factor, $t = (T-T_c)/T_c$ is the reduced temperature, and
$\nu$ is the correlation length critical exponent of the magnetic phase transition.
As we anticipate corrections to scaling to be important in the presence of disorder, we have included the
irrelevant variable $u$ and the corresponding exponent $\delta > 0$. By setting the scale factor
$\lambda=L^{-1}$, we obtain $g(t,L,u) = F(tL^{1/\nu},uL^{-\delta})$ where $F$ is a dimensionless scaling function.
Expanding $F$ in its second argument results in
\begin{equation}
g(t,L,u) = \Phi(tL^{\frac{1}{\nu}}) + uL^{-\delta} \Phi_u(tL^{\frac{1}{\nu}})~.
\end{equation}
If corrections to scaling are negligible ($u=0$), the Binder cumulant vs.\ temperature curves
for different system sizes all cross at the reduced temperature $t=0$ (i.e., at $T=T_c$) and the value $g = \Phi(0)$.

\begin{figure}
\includegraphics[width=\columnwidth]{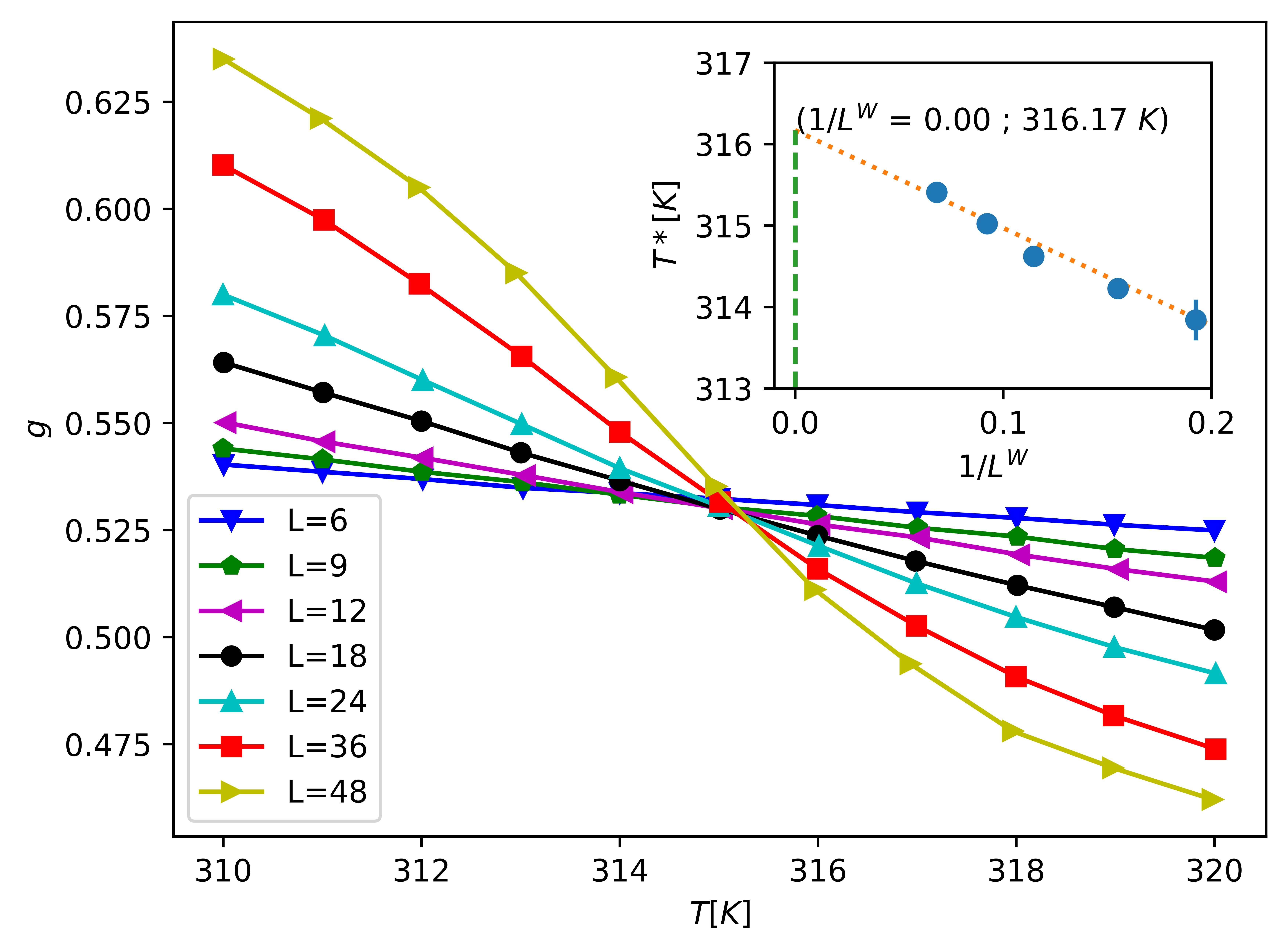}
\caption{Binder cumulant $g$ vs.\ temperature $T$ for several system sizes $L$, dilution $p=0.5$ and weights
$w(12k)=1.25, w(2a)=w(2b)=w(4f_\mathrm{IV})=w(4f_\mathrm{VI})=0.75$. The statistical errors are smaller than the symbol size.
Inset: Extrapolation of the crossing temperature $T^*$ of the $g$ vs.\ $T$ curves for system sizes $L$ and $2L$ to infinite system size,
using eq.\ (\ref{eq:crossingT}) with exponent value $w = 1.5$.}
\label{fig:binder}
\end{figure}

If corrections to scaling cannot be neglected ($u \ne 0$), the analysis is slightly more complicated because
the $g$ vs. $T$ curves do not all cross right at $T_c$. Instead, the crossing point shifts with $L$ and
approaches $t=0$ as $L \rightarrow \infty$.
Expanding the scaling functions $\Phi$ and $\Phi_u$ gives the following expression for
the crossing temperature $T^*(L)$ between the Binder cumulant curves for sizes
$L$ and $cL$ (where $c$ is a constant):
\begin{equation}
T^*(L) = T_c+ b L^{-\omega} \quad \textrm{with} \qquad \omega = \delta+\frac{1}{\nu}
\label{eq:crossingT}
\end{equation}
where $b \sim u$ is a non-universal amplitude \cite{KhairnarLerchVojta21}. In our simulations, we determine the ordering temperature $T_c$ by extrapolating
the numerically found crossing temperatures to infinite system size using the relation (\ref{eq:crossingT}).
An example of this analysis is shown in Fig.\ \ref{fig:binder}.

\section{Results}
\label{sec:results}
\subsection{Critical Ga concentration from percolation theory}
\label{subsec:perc}

Before we turn to the Monte Carlo simulations, we determine the critical Ga concentration, i.e., the Ga concentration
at which the ordering temperature is suppressed to zero, by means of percolation theory. Specifically, we compute
the site percolation threshold of the lattice spanned by the nonfrustrated interactions listed in Table \ref{table:J},
taking into account that the vacancy probability varies from sublattice to sublattice.

We employ a version of the fast percolation algorithm by Newman and Ziff \cite{NewmanZiff01} which allows us
to study systems of up to $300^3$ double unit cells (648 million Fe sites). For each system size, the percolation
threshold is determined from the onset of a spanning cluster, averaged over several hundred disorder configurations.
The results are then extrapolated to infinite system size.

\begin{figure}
\includegraphics[width=\columnwidth]{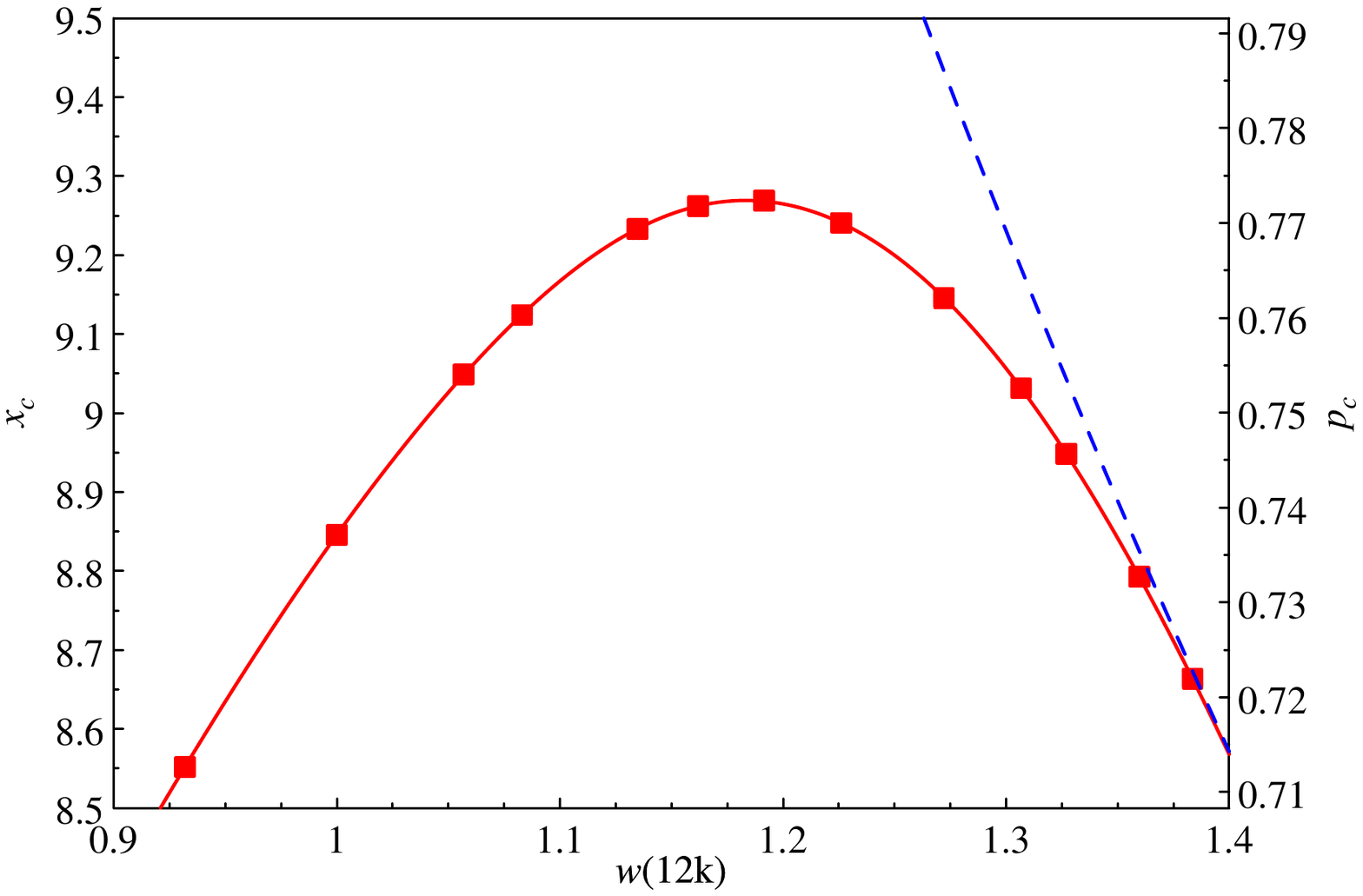}
\caption{Critical Ga concentration $x_c$ and the corresponding critical vacancy probability $p_c=x_c/12$ as functions of the
weight $w(12k)$. The statistical errors of the data points are much smaller than the symbol size. The solid line is a
guide to the eye only. The dashed line represents the Ga concentration $x=12/w(12k)$ at which the $12k$ sublattice becomes
completely depleted of Fe atoms.}
\label{fig:percolation}
\end{figure}

We have applied this analysis to a sequence of systems with varying vacancy weight $w(12k)$ for the $12k$ sublattice.
All other sublattices have identical weights given by Eq.\ (\ref{eq:w12k_only}).
 Figure \ref{fig:percolation} presents the resulting critical Ga concentrations.
For $w(12k) =1$, this calculation reproduces the value $x_c=8.846$ found in Ref.\ \cite{Rowleyetal17}. As $w(12k)$
is increased above unity, the critical Ga concentration first increases. It reaches a maximum of about 9.27 for
$w(12k) \approx 1.18$ before decreasing again. For weights $w(12k) \gtrsim 1.4$, the critical Ga concentration
is given by $x=12/w(12k)$ in very good approximation, indicating that the transition coincides with the complete
depletion of the $12k$ sublattice. It is worth emphasizing that $x_c$ deviates by less than 5\% from its value for
a uniform Ga distribution over a wide range of $w(12k)$ between about 0.9 and 1.4.

\subsection{Magnetic phase boundary}
\label{subsec:PB}

\begin{figure}
\includegraphics[width=\columnwidth]{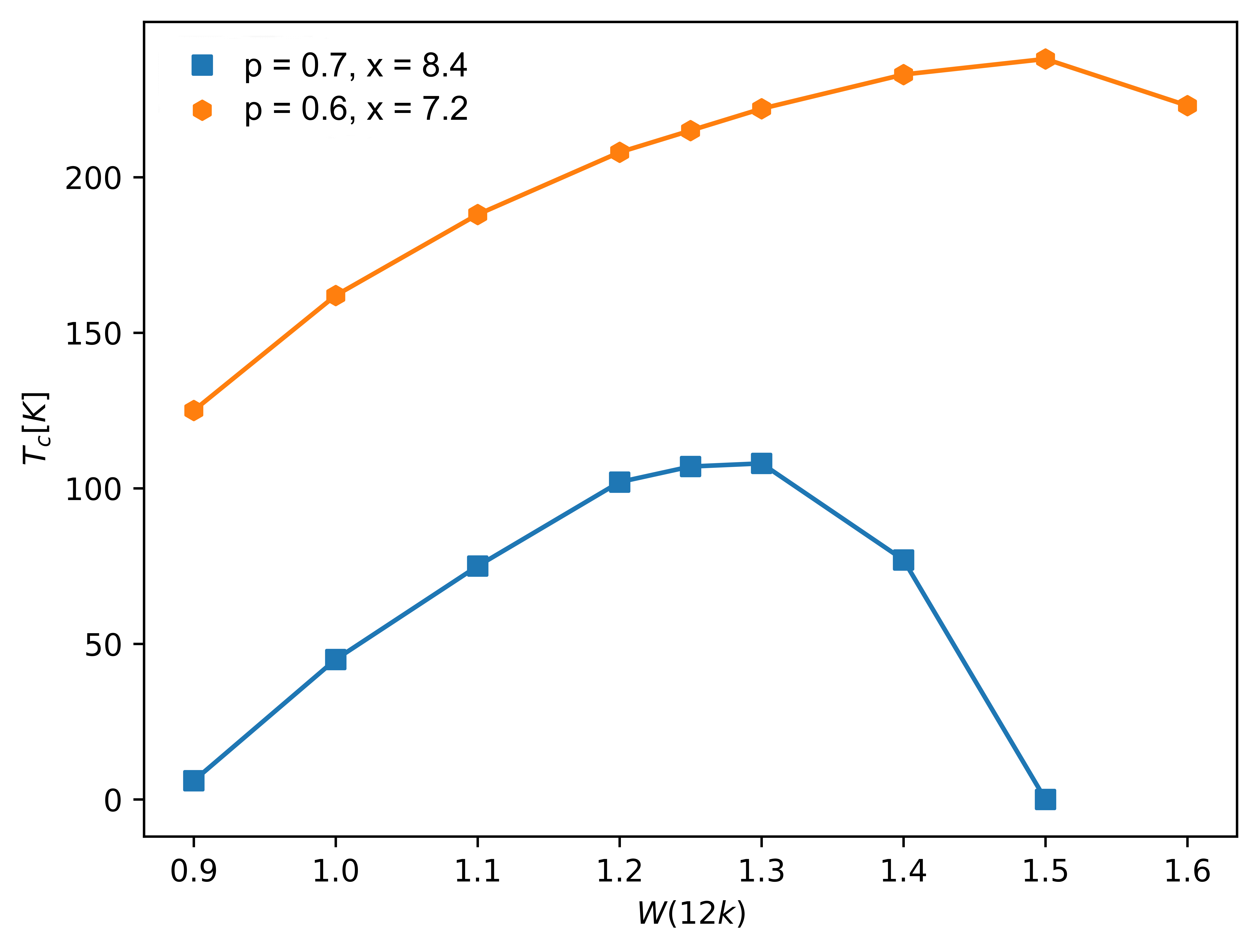}
\caption{Magnetic ordering temperature $T_c$ as a function of the weight $w(12k)$ at fixed Ga concentrations
$x=7.2$ and 8.4. The statistical errors are smaller than the symbol size. The lines are guides to the eye only.
}
\label{fig:Tc_vs_w}
\end{figure}

We now turn to the results of the Monte Carlo simulations. We start by analyzing how the magnetic ordering
temperature $T_c$ depends on the vacancy weights $w$ at fixed overall vacancy concentration. Figure
\ref{fig:Tc_vs_w} presents $T_c$ vs.\ $w(12k)$ at fixed Ga concentration $x=7.2$ and 8.4  (i.e., dilutions $p=0.6$ and 0.7, respectively)
for the same sequence of systems as in Fig.\ \ref{fig:percolation}, i.e. a sequence for which the weight $w(12k)$
differs from all the other weights which are given by Eq.\ (\ref{eq:w12k_only}).
The figure shows that the behavior of $T_c$ is qualitatively similar to that of the
critical Ga concentration $x_c$ discussed in Sec.\ \ref{subsec:perc}. As the weight $w(12k)$ is
increased from the uniform case, $w(12k)=1$, the ordering temperature increases, reaches a maximum, and then
decreases for larger $w(12k)$. However, the relative change of $T_c$ is much stronger than that
of $x_c$. The maximum $T_c$ is about 50\% higher than the $T_c$ value in the uniform case for $x=7.2$
and twice as high for $x=8.4$.
Moreover, for weights $w(12k)$ in the range from 1.2 to 1.3, the ordering temperature roughly agrees
with the experimental value obtained from the data in Refs.\ \cite{Rowleyetal17,ALWD02}
(see Fig.\ \ref{fig:experiment}).

\begin{figure}
\includegraphics[width=\columnwidth]{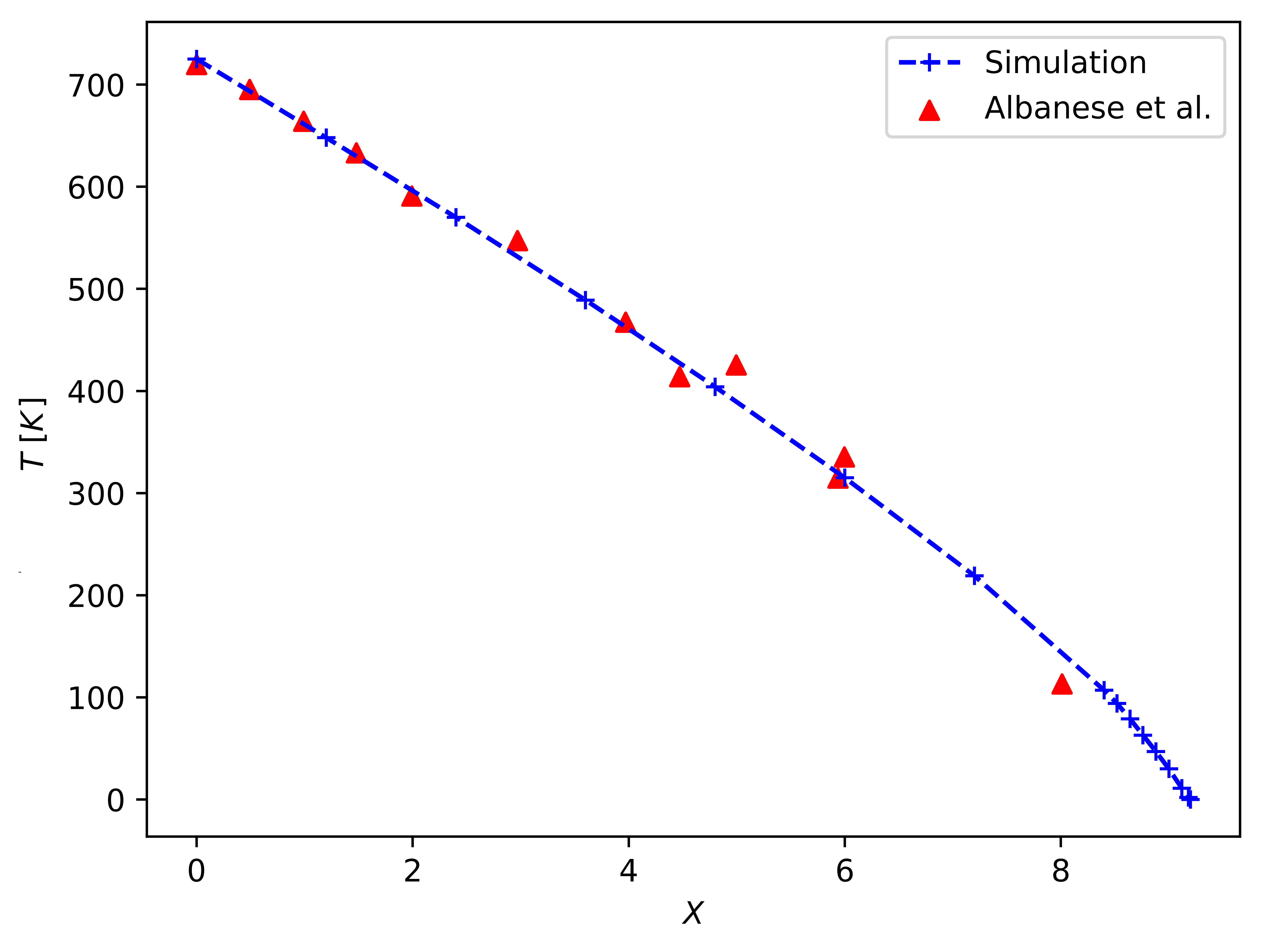}
\caption{Magnetic ordering temperature $T_c$ vs.\ Ga concentration $x$ for $w(12k)=1.25$.
All other weights take the value $2-w(12k)$. The statistical errors are smaller than the symbol size.
The dashed line is a guide to the eye only.
The experimental values stem from Refs.\ \cite{Rowleyetal17,ALWD02}.}
\label{fig:phaseboundary}
\end{figure}

Based on this observation, we have computed the ordering temperature $T_c(x)$ over the entire $x$-range between
0 and $x_c$ for the weight $w(12k)=1.25$. The resulting phase boundary for $w(12k)=1.25$ is shown
in Fig.\ \ref{fig:phaseboundary} together with the corresponding experimental data.
Clearly, the Monte Carlo results for $w(12k)=1.25$ are in excellent agreement with experiment, in contrast
to the Monte Carlo results for the uniform case, $w(12k)=1$ shown in Fig.\ \ref{fig:experiment}.
The corresponding $T_c(x)$ curves for weights $w(12k)=1.2$ and 1.3 deviate only slightly from the phase boundary for
$w(12k)=1.25$.

For comparison, we have also analyzed a case in which the impurity weights in both the $12k$ sublattice
and the $2a$ sublattice are increased, $w(12k)=w(2a)=1.2$, whereas the other weights are reduced,
$w(2b)=w(4f_\mathrm{IV})=w(4f_\mathrm{VI})=0.72$. The resulting phase boundary is virtually indistinguishable
from that for the case (\ref{eq:w12k_only}) with $w(12k)=1.2$.

\subsection{Saturation magnetization}
\label{subsec:Ms}

\begin{figure}
\includegraphics[width=\columnwidth]{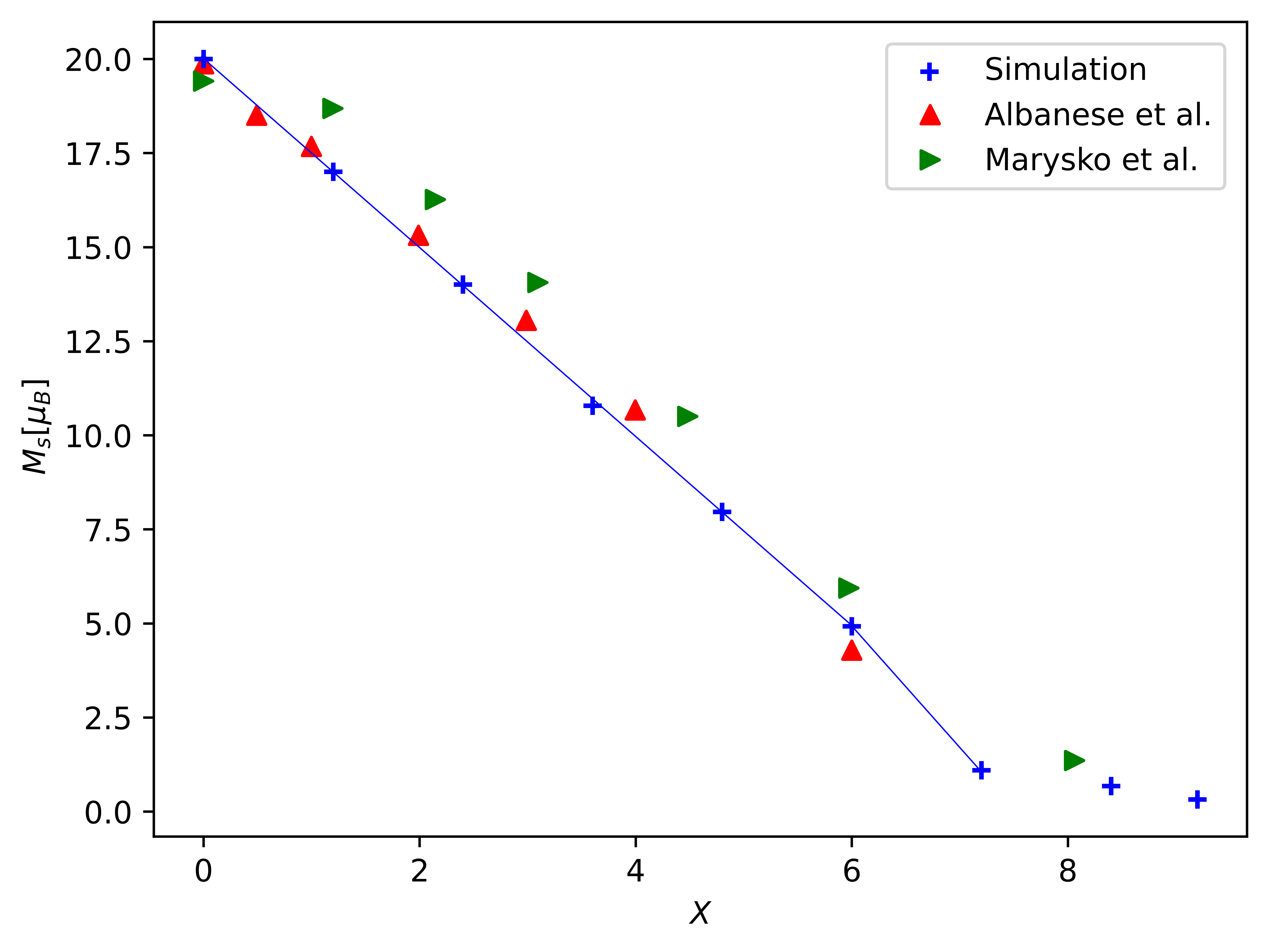}
\caption{Low-temperature saturation magnetization $M$ (in $\mu_B$ per formula unit) vs.\ Ga concentration $x$
for $w(12k)=1.25$. All other weights take the value $2-w(12k)$. The line is a guide to the eye only.
The experimental values stem from Refs.\ \cite{MaryskoFraitKrupicka97,ALWD02}.
}
\label{fig:Ms}
\end{figure}

In addition to the ordering temperature, we have also calculated the low-temperature limit of the
magnetization $\mathbf{M}$. As all interactions in our model Hamiltonian are nonfrustrated, this
value can be directly compared to the low-temperature saturation magnetization $M_s$ measured in experiment.
Figure \ref{fig:Ms} presents $M_s$ vs.\ $x$ for $w(12k)=1.25$  [and all other weights given
by Eq.\ (\ref{eq:w12k_only})] together with the experimental values from Refs.\
\cite{MaryskoFraitKrupicka97,ALWD02}.
The figure demonstrates that the increased $w(12k)$ weight leads to a reduction of the saturation magnetization
compared to the case of uniform vacancy distribution and produces a good agreement between our model and the
experimental data.

The fact that an increased vacancy weight $w(12k)$ leads to an \emph{increase} of $T_c$ and $x_c$ but
a \emph{decrease} of $M_s$ appears somewhat counterintuitive at first glance. However, it can be readily
explained by the ferrimagnetic order in the hexaferrites. An increased $w(12k)$ weight reduces the number
of Fe atoms in the majority (spin-up) sublattices whereas it increases the number of Fe atoms in the
minority (spin-down) sublattices. As a result, the difference between the numbers of spin-up and spin-down
Fe ions is reduced, leading to a significant reduction of $M_s$. Note that this explanation does not rely on
non-collinear magnetic order caused by the subleading (frustrating) interactions. We will return to this point in the
concluding section.

\subsection{Sublattice magnetizations}
\label{subsec:sublattice}

\begin{figure}
\includegraphics[width=\columnwidth]{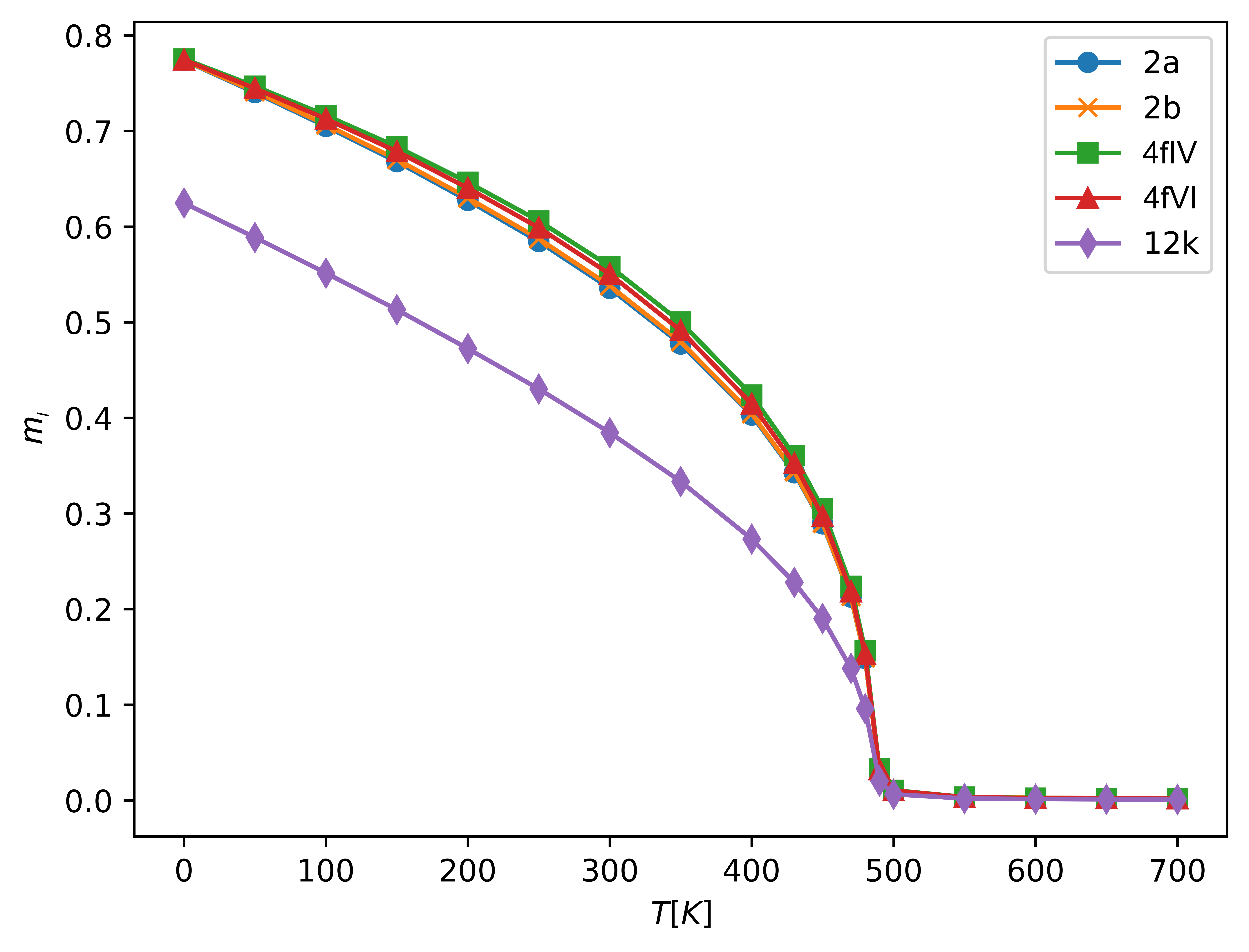}
\caption{Sublattice magnetizations $m_l$ vs.\ temperature $T$ for $x=3.6$ and $w(12k)=1.25$.
All other weights take the value $2-w(12k)$.
}
\label{fig:ml}
\end{figure}

In addition to the ferrimagnetic order parameter $\bm{\psi}$ and the total magnetization $\mathbf{m}$, we have also
calculated how the sublattice magnetizations $\mathbf{m}_l$ [defined in Eq.\ (\ref{eq:ml})]
depend on the impurity concentration $x$, the weights $w$ and the temperature $T$.
Figure \ref{fig:ml} shows the sublattice magnetizations as functions of temperature at fixed
$x=3.6$ and $w(12k)=1.25$. [All other weights are equal and given by Eq.\ (\ref{eq:w12k_only}).]
As expected, $m_l(12k)$ is significantly lower than all the other sublattice magnetizations because
the vacancy concentration in the $12k$ sublattice is higher than those of the other sublattices. In fact,
the zero-temperature limit of $m_l$ agrees with the fraction of occupied sites in each sublattice.
The differences of the sublattices magnetizations between the other sublattices ($2a$, $2b$, $4f_\mathrm{IV}$,
$4f_\mathrm{VI}$) are very small. They reflect the differences in the environments of the Fe atoms
in the different sublattices.

\subsection{Critical behavior at $x_c$}
\label{subsec:critical}

Finally, we turn to the critical behavior of the zero-temperature phase transition at $x_c$. To this end,
we analyze the the dependence of $T_c$ on the distance $x-x_c$ from the zero temperature transition.
For the case of a uniform vacancy distribution, Khairnar et al.\ \cite{KhairnarLerchVojta21} showed
that $T_c \sim (x_c-x)^\phi$ with $\phi=1.12$ in a narrow asymptotic region close to $x_c$. This value
of the crossover exponent $\phi$ agrees with the predictions of classical percolation theory
\cite{StaufferAharony_book91,ShenderShklovskii75,Coniglio81}, confirming that transition at $x_c$ is a percolation transition.
The pre-asymptotic behavior of $T_c$ further away from $x_c$
still followed a power law in $(x_c-x)$, but with a nonuniversal crossover exponent.
Its value was below unity but well above the experimentally observed 2/3.

Here we employ the same analysis for the case of a nonuniform vacancy distribution, specifically for
our sequence of systems with increased vacancy weight $w(12k)$ and reduced weights (\ref{eq:w12k_only})
for all other sublattices. We find two different regimes, depending on $w(12k)$.

\begin{figure}
\includegraphics[width=\columnwidth]{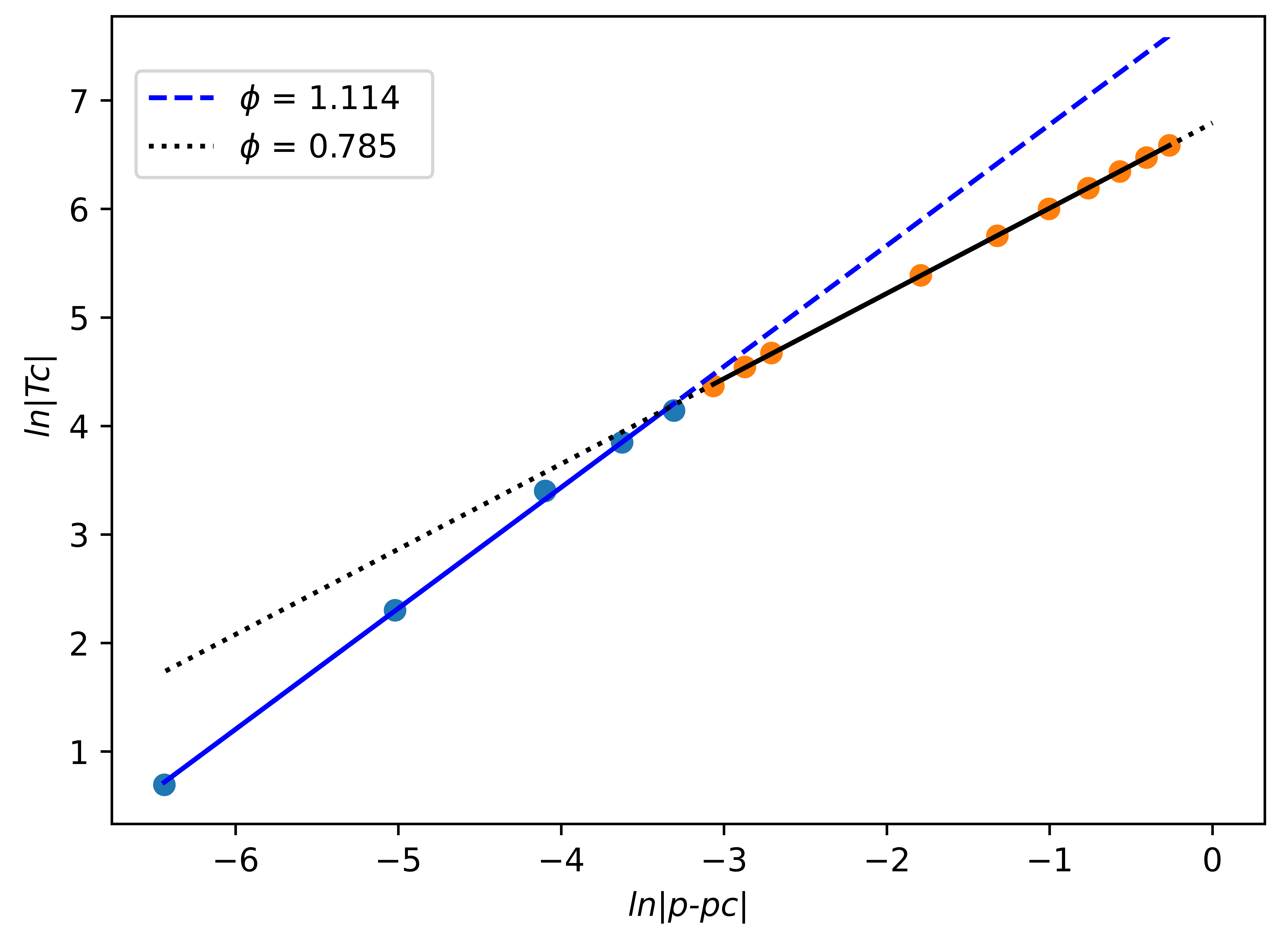}
\caption{Ordering temperature $T_c$ vs.\ distance from the percolation threshold $p_c-p$ for
 $w(12k)=1.25$. All other weights take the value $2-w(12k)$. The dashed line is a power-law fit
$T_c \sim (p_c-p)^\phi$ of the data points close to $x_c$. The dotted line is the corresponding fit of
the preasymptotic behavior. The statistical errors are about a symbol size for the data closest to $p_c$
and smaller further away from $x_c$.}
\label{fig:critical1}
\end{figure}

In the first regime, the zero-temperature transition occurs (as a function of increasing $x$) before
the $12k$ sublattice is completely depleted of Fe atoms. As shown in Fig.\ \ref{fig:percolation},
this happens for $w(12k) \lessapprox 1.4$. In this regime the behavior of $T_c$ with $x-x_c$ is analogous
to the case of uniform dilution: Asymptotically close to $x_c$, the ordering temperature follows
$T_c \sim (x_c-x)^\phi$ with the $\phi$ value from percolation theory. Further away from $x_c$, the
behavior crosses over to a weaker dependence on $x$. This is illustrated in Fig.\ \ref{fig:critical1}
for $w(12k)=1.25$.

\begin{figure}
\includegraphics[width=\columnwidth]{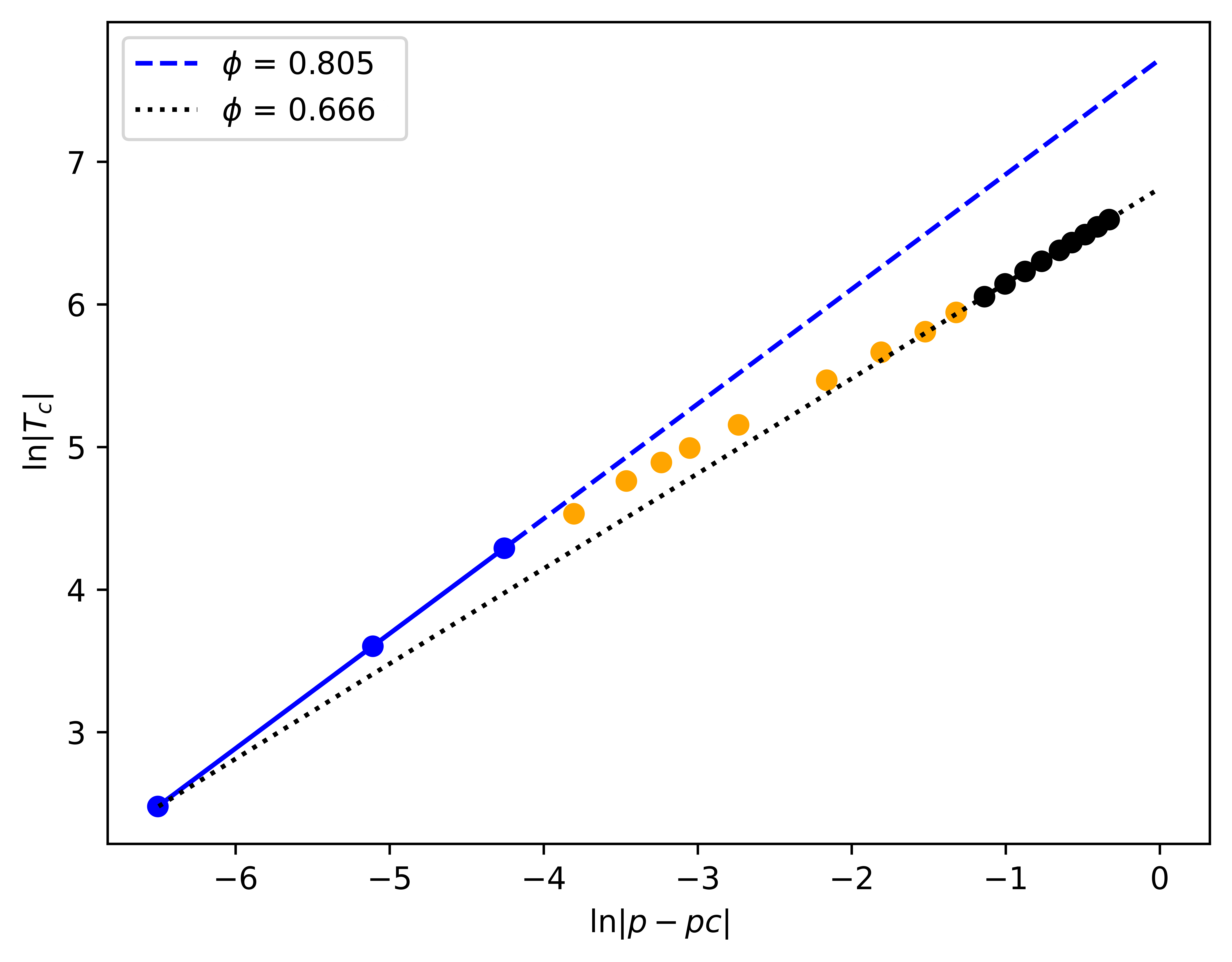}
\caption{Ordering temperature $T_c$ vs.\ distance from the percolation threshold $p_c-p$ for
 $w(12k)=1.5$. All other weights take the value $2-w(12k)$. }
\label{fig:critical2}
\end{figure}

In the second regime, the zero-temperature transition coincides with the complete depletion
of the $12k$ sublattice of Fe atoms, as is the case for $w(12k) \gtrapprox 1.4$ (at least in very good approximation).
Figure \ref{fig:critical2} shows the dependence of $T_c$ on $x_c-x$ for $w(12k)=1.5$ with $x_c=12/w(12k)=8$.
Clearly, the shape of the phase boundary $T_c(x)$ differs from the percolation scenario described
above, suggesting a different universality class of the zero-temperature critical point.

\section{Conclusions}
\label{sec:conclusions}

In summary, motivated by the disagreement between experimental data and theoretical predictions, we have
revisited the magnetic properties of diluted hexagonal ferrites, in particular PbFe$_{12-x}$Ga$_x$O$_{19}$.
We have focused on the effects of an uneven distribution of the nonmagnetic Ga impurities over the five
distinct Fe sublattices.

Our ab-initio density-functional calculations have shown that the preferred sublattice for the Ga atoms is the
$12k$ sublattice. To include this preference in the Monte-Carlo simulations, we have created model impurity
distributions with increased vacancy probability in the 12k sublattice and correspondingly reduced probabilities
in all other sublattices. These probabilities are parameterized by weights $w$ such that the vacancy probability
in sublattice $Y$ is given by $w(Y) \, p$ with $p=x/12$ the overall vacancy probability.

For appropriately chosen sublattice weights [$w(12k) \approx 1.2$ to 1.3 and the other $w$ correspondingly
reduced], our Monte Carlo results for the phase boundary $T_c(x)$ and the low-temperature
saturation magnetization $M_s$ are in excellent agreement with the experimental data of Refs.\
\cite{MaryskoFraitKrupicka97,ALWD02,Rowleyetal17}. This indicates that the uneven distribution of the
Ga impurities is the main reason for the discrepancies between the measurements and previous theoretical
work in the literature. Notably, the uneven Ga distribution explains why the experimental saturation
magnetization drops rapidly with increasing $x$ whereas the critical temperature $T_c$
decreases much more slowly.

We have also studied the critical behavior of the zero-temperature phase transition at the critical
Ga concentration $x_c$. If this transition happens before any of the sublattices becomes completely
depleted of Fe, the critical behavior of $T_c$ follows the predictions of percolation theory.
For a more unequal Ga distribution (larger weight $w(12k) \gtrapprox 1.4$), the zero-temperature transition
coincides with the emptying of the $12k$ sublattice. This leads to a different critical behavior of
$T_c(x)$.

Our results suggest that the agreement between the experimental $T_c$ data and the
striking 2/3 power-law behavior of the phase boundary $T_c(x)$ put forward in Ref.\
\cite{Rowleyetal17} is actually ``accidental.'' It appears to be the result of the particular Ga weights
rather than a fundamental principle. In fact, the data in Fig.\ \ref{fig:phaseboundary} suggest that the
2/3 power law may not accurately describe the data close to $x_c$. Testing this predictions
requires additional experiments at Ga concentrations close to $x_c$.

The Ga distributions employed in our simulations should be considered simple models rather than a quantitatively
accurate description of the real materials. For example, we do not include possible dependencies of the weights $w$ on the
concentration $x$, and we neglect any correlations between neighboring impurities. Extracting a more detailed
description of the Ga distribution from our Monte Carlo simulations requires additional experimental
input beyond $T_c$ and $M_s$. In particular, fully disentangling the Ga concentrations in all five sublattices
would require the measurement of sublattice magnetization curves (or equivalent local information) over the full $x$
range. To the best of our knowledge, such experimental data are not yet available.

Our present simulations do not include any of the subleading exchange interactions that frustrate the
ferrimagnetic order. As a consequence, the magnetic order remains collinear over the entire $x$ range.
Simulations performed in Ref.\ \cite{KhairnarLerchVojta21} showed that subleading frustrating interactions
and the resulting non-collinear order lead to a reduction in $T_c$. This implies that
the frustrating interactions (by themselves) cannot explain the disagreement between the experiments
and previous theoretical work. Even though our results demonstrate that non-collinear order is not
the main reason for the rapid drop of the saturation magnetization $M_s$ with $x$, the frustrating
interactions and the resulting non-collinear order likely become important at larger $x$ close to the
zero-temperature transition.

We hope that our study will encourage further experimental and theoretical work that helps resolving
the open questions about the behavior of PbFe$_{12-x}$Ga$_x$O$_{19}$ and other diluted hexaferrites.\\

\acknowledgments

This work has been supported in part by the National Science Foundation under
Grant Nos.\  DMR-1828489 and OAC-1919789. The simulations were performed on the
Pegasus and Foundry clusters at Missouri S\&T. We acknowledge helpful discussions with Stephen Rowley.

\appendix
\section{Details of the ab-initio calculations}
\label{sec:ab-initio}

To determine preferred distribution of Ga atoms among the five distinct Fe sublattices in the hexaferrite,
the electronic and magnetic properties of Ga-doped PbFe$_{12}$O$_{19}$ were studied using first-principles density
functional calculations as implemented in the Vienna Ab-initio Simulation Package (VASP) \cite{KresseHafner93,KresseHafner94,KresseFurthmuller96,KresseFurthmuller96b}.
The generalized gradient approximation (GGA) in the Perdew-Burke-Ernzerhof (PBE) form \cite{PerdewBurkeErnzerhof96,PerdewBurkeErnzerhof97}
within the projector augmented-wave method was used \cite{Blochl94,KresseJoubert99}. The GGA+U method was employed with an on-site
Coulomb $U=0$ and $U=3$\,eV for Fe--$d$-states.

It was found that the $U$-correction is necessary to attain
the correct antiferromagnetic ordering in PbFe$_{12}$O$_{19}$: the calculated magnetic moments are found to be
$4.10\,\mu_B$, $4.00\,\mu_B$, $-3.96\,\mu_B$, $-4.01\,\mu_B$, and $4.12\,\mu_B$ for Fe in octahedral 2a, pseudo-hexahedral 2b,
tetrahedral 4f$_{IV}$, octahedral 4f$_{VI}$, and octahedral 12k position, respectively, when $U=3$\,eV, whereas at $U=0$ the moments
are $2.28\,\mu_B$, $3.72\,\mu_B$, $-3.47\,\mu_B$, $0.81\,\mu_B$, and $3.45\,\mu_B$. Further increase in U value ($U=4$, 5, or 6\,eV) lead to an
insignificant increase in the Fe magnetic moments with no effect on the antiferromagnetic ordering. Therefore, $U=3$\,eV  was employed
in this work to study the effect of Ga on the electronic and magnetic properties of PbFe$_{12}$O$_{19}$. The conventional,
double unit cell, Pb$_{2}$Fe$_{24}$O$_{38}$, was used for both undoped and Ga-doped structures. Both the internal
atomic positions and the lattice parameters in all structures were optimized using force and total energy minimization until the
Hellmann-Feynman force on each atom was below $0.01$\,eV/\r{A}.
Brillouin-zone sampling was done with $\Gamma$-centered Monkhorst-pack with $k$-mesh of at least 8x8x2; the cut-off energy of 500\,eV and
 Gaussian smearing were used.

\bibliography{../00Bibtex/rareregions}

\end{document}